\documentclass[aps,prl,amsmath,amssymb,twocolumn,floatfix,10pt,showpacs]{revtex4-1}
\usepackage{graphicx}
\usepackage{hyperref}
\usepackage{color}

\newcommand{\bs}{\;\;\;\;\;}

\newcommand{\ve}{\mathbf}

\newcommand{\D}{{\rm d}}

\begin{document}

\title{Strong correlations at topological insulator surfaces and the breakdown of the bulk-boundary correspondence}
\author{Manuel J. Schmidt}
\affiliation{Institut f\"ur Theoretische Festk\"orperphysik, RWTH Aachen University, 52056 Aachen, Germany}
\date{\today}
\pacs{75.70.Rf, 73.43.Nq, 73.20.-r}


\begin{abstract}
The criteria for strong correlations on surfaces of three-dimensional topological insulators are discussed. Usually, the Coulomb repulsion at such surfaces is too weak for driving a phase transition to a strongly correlated regime. I discuss a mechanism and possibilities of its experimental implementation by which the strength of the Coulomb interaction can be tuned over a wide range. In the strongly interacting regime, the surface states are gapped, even though the topological classification of the bulk band structure predicts gapless surface states.
\end{abstract}

\maketitle

The field of topological insulators (TI) started a few years ago with the investigation of spin-orbit interactions (SOI) in graphene \cite{kane_mele_QSHE_2005}, where it was shown that graphene with an artificially large SOI is insulating in the bulk, but has perfectly conducting channels at its boundaries. It was quickly noticed \cite{sheng_qshe_topology_2006,qi_bulk_edge_correspondence_2006} that the presence of these boundary states is intimately connected to the topological properties of the bulk material (bulk-boundary correspondence); a topological classification of the insulating band structure - a bulk property - gives rise to two distinct material classes: trivial and nontrivial insulators. The latter must have perfectly conducting boundary channels which are robust against all time-reversal invariant single-particle perturbations \cite{ti_surface_state_robustness}. In this early stage of topological classifications the electron-electron interaction played essentially no role and the theory was based on band structure arguments.

More recently, however, researchers noticed that the topological classification of insulators may be extended to include the Coulomb repulsion \cite{wang_top_class_interaction_2010}. The possibility of performing unbiased and numerically exact quantum Monte-Carlo simulations of the Kane-Mele-Hubbard model \cite{thomas} - a paradigmatic model for a strongly interacting two-dimensional (2D) TI - ignited much research exploring the interplay of interactions and topology. The topological bulk properties can be destroyed by a strong Coulomb repulsion, which may drive a quantum phase transition to a time-reversal symmetry-broken state \cite{thomas}. But also the direct effect of interactions on the helical edge states has been studied \cite{hohenadler_corr_edges_2012}. It turned out that, while the helical edge states are robust against weak interactions \cite{stability_qshe}, strong interactions lead to dramatic changes of the single-particle properties near the edges \cite{hohenadler_corr_edges_2012}.

The investigation of correlations in TIs were largely restricted to 2D TIs with 1D states at their edges. Strong correlations at surfaces of 3D TIs have not yet received much attention. This might be due to the rather large interaction strengths that are usually needed for driving a phase transition of those surfaces. In this Letter, I show how the surface states can be tuned to a strongly interacting regime, while the bulk stays practically non-interacting. The mechanism is based on the modification of the non-interacting spectrum of the surface states within the bounds set by the topological bulk properties. The modification must be done in such a way that the local density of states is enhanced near the Dirac point. Such modifications will be shown to be accomplishable via changing the material properties close to the surface, for instance by growing a thin layer of a different TI on top of a given TI surface by molecular beam epitaxy. 

{\it Relation of bulk and edge interactions.} In TIs there are two energy scales competing with the Coulomb repulsion, modeled by a Hubbard interaction $U$ in this work. In the insulating bulk this is the energy gap $\Delta$. For $\Delta\lesssim U$ the bulk may enter a time-reversal symmetry-broken phase in which the original topological classification becomes meaningless. On the other hand, the gapless surface states become correlated if $U\gg \hbar v_F/a$, with $v_F$ the Fermi velocity of the surface Dirac cone and $a$ the lattice constant ($a=1$ in the following). This work focusses on the regime of an effectively non-interacting bulk, with a strongly interacting surface. Usually, however, the bulk gap $\Delta\propto \hbar v_F/a$ is proportional to the surface Fermi velocity $v_F$, because the surface states connect valence and conduction band. It is not obvious that a regime of a weakly interacting bulk and a strongly interacting surface (i.e., $\Delta\gg U\gg \hbar v_F/a$) is reachable.

The above argument for $\Delta \propto v_F$ relies on the assumption that the linear term in the surface spectrum is dominant. This, however, is not enforced by the nontrivial topology. The much discussed surface Dirac cones are only the leading terms in the expansion of the surface states' energy spectrum. In general there are also higher order terms and the corresponding Hamiltonian of the surface states is of the form
\begin{equation}
H_{\rm sur} = (v_F  + g_3 \ve k^2 + g_5 \ve k^4 + ...) (k_x s_x + k_y s_y ),
\end{equation}
with $\ve k=(k_x,k_y)$ the surface momentum and $s_\mu$ the spin Pauli matrices. Note that there may also be more complicated $k^3$ terms breaking the rotational symmetry \cite{liu_model_hamiltonian_2010}. Those terms are not considered in this work as they do not affect the general idea discussed here.

Assuming that the surface states exist for $|\ve k|<\Lambda \sim 1/a$, the bulk gap $\Delta\simeq  v_F\Lambda + g_3 \Lambda^3 + O(\Lambda^5)$. This shows that $v_F$ can in principle be varied independently from the bulk gap. In the remainder of this work the bulk gap $\Delta$ is assumed to be fixed so that $g_3 = \Delta \Lambda^{-3} -  v_F \Lambda^{-2}$ is a function of $v_F$. How these cubic terms may be manipulated in actual materials will be discussed below.

{\it Interaction effects.} The Coulomb repulsion in the surface states is modeled by a Hubbard-like interaction
\begin{equation}
H_1 = \frac{U}{L^2} \sum_{\ve k,\ve k',\ve q} \Gamma(\ve k,\ve k',\ve q) c^\dagger_{\ve k+\ve q\uparrow} c_{\ve k\uparrow} c^\dagger_{\ve k'-\ve q\downarrow} c_{\ve k'\downarrow},
\end{equation}
where the fermionic operator $c_{\ve k\sigma}$ annihilates an electron with spin $\sigma$ and momentum $\ve k$ in a surface state. $L^2\rightarrow\infty$ is the surface area. The interaction vertex $\Gamma(\ve k,\ve k',\ve q)$ encodes the localization properties of the surface states in the direction perpendicular to the TI surface. $H_1$ is inspired by the effective model for interacting graphene edge states \cite{schmidt_tem_2010}. The exact form of the vertex function $\Gamma$ depends on the details of the TI surface, but its most important and generic feature is that it reduces to zero if one of the momenta $\ve k+\ve q,\ve k,\ve k'-\ve q,\ve k'$ approaches the ultraviolet cutoff $\Lambda$, at which the surface states merge into the bulk bands. At this momentum, the transverse wave function is completely delocalized into the bulk. This strong momentum dependence of the interaction vertex is a general feature of topological confinement. Following Ref. \onlinecite{luitz_ed_2011}, $\Gamma$ is modeled by
\begin{equation}
\Gamma(\ve k,\ve k',\ve q) = \mathcal N(\ve k+\ve q)\mathcal N(\ve k)\mathcal N(\ve k'-\ve q)\mathcal N(\ve k'),
\end{equation}
with $\mathcal N(\ve k) = \sqrt{1-|\ve k|/\Lambda}$ playing the role of the momentum-dependent wave function weight at the surface. The larger $\mathcal N(\ve k)$ the more localized is the surface state in the direction perpendicular to the surface; and more localization means stronger Coulomb repulsion. Note also that due to the fermion doubling theorem, the theory of one single TI surface cannot be formulated on a real space lattice.

In this work the interaction $H_1$ is treated in mean-field approximation. Since the SU(2) symmetry is broken by the SOI, it is most important to use an SU(2) invariant mean-field decoupling of $H_1$, i.e., 
\begin{equation}
H_1^{\rm MF} = \frac U4 \ve m^2 - \frac U2 \sum_{\ve k\mu\sigma\sigma'} \mathcal N^2(\ve k) m_\mu c^\dagger_{\ve k\sigma} s^\mu_{\sigma\sigma'}c_{\ve k\sigma'}
\end{equation}
with the components of the spin polarization $\ve m$
\begin{equation}
m_\mu = \frac1{L^2} \sum_{\ve k\sigma\sigma'}\mathcal N^2(\ve k) \langle c^\dagger_{\ve k\sigma} s^\mu_{\sigma\sigma'}c_{\ve k\sigma'}\rangle,\bs \mu=x,y,z .
\end{equation}

It is most convenient to formulate this mean-field approximation in terms of a variational principle. The total ground state energy of the interacting surface states depends on the spin polarization $\ve m$ and is given by
\begin{multline}
E_{\rm tot}(\ve m) = \frac U4\ve m^2 \\+ \sum_{s=\pm1}\int \frac{ \D^2\ve k}{(2\pi)^2} \epsilon_s(\ve k,\ve m) \Theta[\epsilon_F(\ve m) - \epsilon_s(\ve k,\ve m)] \label{toten}
\end{multline}
with the single-particle energies
\begin{equation}
\epsilon_\pm(\ve k,\ve m) =\pm\biggl[ \sum_{\mu=x,y,z} (\gamma_\mu - h_\mu)^2\biggr]^{1/2}
\end{equation}
and $\gamma_{x,y}= ( v_F + g_3 k^2) k_{x,y}$, $\gamma_z=0$ and $h_\mu = U m_\mu \mathcal N^2(\ve k)/2$. The Fermi energy $\epsilon_F(\ve m)$ is fixed by requiring a constant number of electrons $N_e$ in the $N$ surface states. The fraction of occupied states $N_e/N=(1+f)/2$ is characterized by the filling factor $f$. For half-filling, $f=0$. For $f=1$ ($f=0$) all surface states are filled (empty). The integral (\ref{toten}) cannot be solved exactly for general $\ve m$, but the expansion of $E_{\rm tot}(\ve m)$ at $\ve m=0$ is accessible. In the following, the two opposite limits $g_3\rightarrow0$ (normal Dirac cone) and $v_F\rightarrow0$ (Dirac hourglass) are discussed. 

The critical interaction strength $U_c$, above which a finite spin polarization $\ve m\neq 0$ is favored, is calculated by solving $\partial_{m_\mu}^2  E_{\rm tot}(\ve m) =0$. Because of the broken spin-rotation symmetry [SU(2)], $U_c$ depends on whether $\ve m$ is in the xy plane ($U_{c,xy}$) or in z direction ($U_{c,z}$). If $U_{c,z}<U_{c,xy}$ there is a second order phase transition at $U=U_{c,z}$ to a spin-polarized phase with Ising anisotropy $m_z\neq 0$. In this phase the bulk-boundary correspondence breaks down: the surface Dirac cone is gapped although the bulk band structure predicts gapless surface states.

\begin{figure}[!ht]
\centering
\includegraphics[width=200pt]{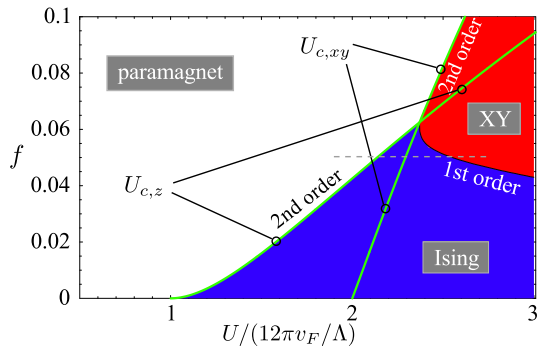}
\caption{(Color online) Phase diagram of the normal Dirac cone. The green curves are the analytical formulas for the critical lines [Eq. (\ref{critline_cone})]. The order of the phase transitions are indicated. One movie in the supplemental material \cite{supp} shows the ground state energies along the cut indicated by the dashed grey line.}
\label{fig_phases}
\end{figure}

For the normal Dirac cone $(g_3=0$), the critical interaction strengths for xy- and z-polarization are
\begin{align}
U_{c,xy} &= \frac{1}\Lambda \frac{24\pi v_F}{1-3f+2 f^{3/2}}, &
U_{c,z} &= \frac{1}\Lambda \frac{12\pi v_F}{(1-\sqrt f)^3},\label{critline_cone}
\end{align}
respectively. For half-filling ($f=0$), $U_{c,xy}>U_{c,z}$ so that the magnetic state has an Ising anisotropy. For larger fillings the $xy$ instability becomes dominant. Thus, there is a critical filling $f_c=1/16$ above which the correlated Dirac cone is easy plane. Figure \ref{fig_phases}(a) shows the phase diagram of the normal Dirac cone. The 2nd order phase boundaries are complemented by a 1st order transition line between the $xy$ and the Ising phase.

For the normal Dirac cone without cubic terms in the spectrum, the critical interactions have an explicit UV cutoff ($\Lambda$) dependence. This is a rather general feature of Dirac physics and is also frequently observed in the context of graphene (see, e.g., Refs. \onlinecite{graphene_susceptibility,jelena_prb}). The presence of the UV cutoff may be traced back to the linear energy dependence of the density of states.

\begin{figure}[!ht]
\centering
\includegraphics[width=220pt]{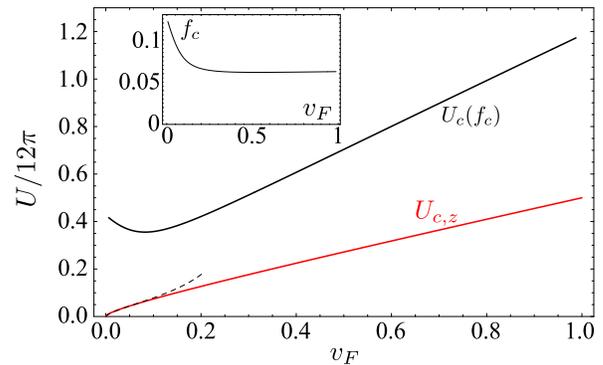}
\caption{Suppression of the critical interaction and stabilization of the Ising phase by lowering $v_F$: $U_{c,z}$ is shown for half-filling ($f=0$). The curve labeled by $U_c(f_c)$ shows the interaction strength at which the transition to the XY phase occurs. The inset shows the critical filling $f_c$ above which the spin polarization is in-plane. The dashed line shows the approximation Eq. (\ref{ucz_approx}). $\Lambda=1$ and $\Delta=1$ in this plot.}
\label{fig_vFplot}
\end{figure}

\begin{figure}[!ht]
\centering
\includegraphics[width=230pt]{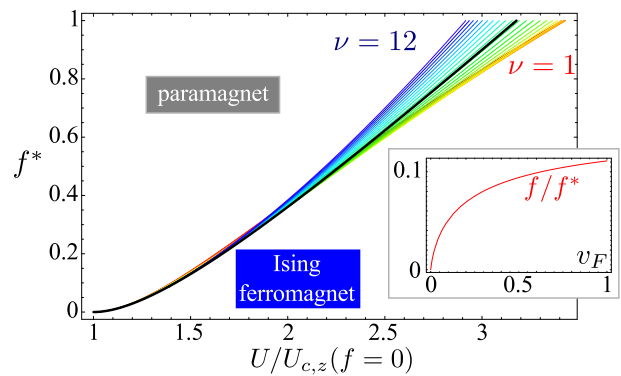}
\caption{(Color online) Phase boundary between the paramagnet and the Ising ferromagnet phase for different $v_F=2^{-\nu}$ with $\nu$ from 1 (red) to 12 (blue), thus varying over three orders of magnitude. The thick black line is the approximate scaling form Eq. (\ref{scalingform}). The inset shows how the real filling $f$ decreases for $v_F\rightarrow0$. $\Lambda=1$ and $\Delta=1$ in this plot.}
\label{fig_scalingform}
\end{figure}

For the Dirac hourglass ($g_3\Lambda^2 \gg v_F$), the exact formulas for the $U_{c,z}$ and $U_{c,xy}$ are accessible but rather involved \cite{supp}. The critical filling $f_c$, at which the XY phase sets in, is shown in the inset of Fig. \ref{fig_vFplot}. It remains finite for any $v_F$ and the corresponding critical interaction $U_c(f_c) = U_{c,z}(f_c) = U_{c,xy}(f_c)$ is large for all $v_F$. Thus in the experimentally relevant limits of small $U,v_F,f$ the Ising instability is dominant so that we may focus on $U_{c,z}$. The exact formula for $U_{c,z}$ for all $v_F$ is plotted for half-filling in Fig. \ref{fig_vFplot}. It vanishes for $v_F\rightarrow0$ and in this limit
\begin{equation}
U_{c,z}(f=0) \simeq \frac{8\sqrt{v_Fg_3}}{1-v_F/g_3\Lambda^2+\sqrt{v_F/g_3\Lambda^2}\log(v_F/g_3\Lambda^2)} \label{ucz_approx}
\end{equation}
is an excellent approximation (dashed line in Fig. \ref{fig_vFplot}). This is one of the main results of this work. For arbitrary filling $f>0$ the critical point approximately follows a scaling form
\begin{equation}
U_{c,z}(f) \simeq U_{c,z}(f=0) (1+\sqrt{f^*} + f^*) + O(f^*)^{3/2} \label{scalingform}
\end{equation}
with the renormalized filling $f^* = (4\pi v_F k_F /U_{c,z}(f=0))^2$. As shown in Fig. \ref{fig_scalingform}, Eq. (\ref{scalingform}) is a good approximation to the true phase boundary. It should be noted that, unlike in the pure Dirac case ($g_3=0$), $\Lambda\rightarrow\infty$ is in principle a well-behaved limit for the Dirac hourglass; the $k^3$ terms regularize the UV divergence in the integrals. Nevertheless, the logarithmic corrections in Eq. (\ref{ucz_approx}) are still significant down to $v_F/g_3\Lambda^2 \sim 10^{-3}$.

{\it Properties of the magnetic state.} The expansion of $E_{\rm tot}(\ve m)$ up to 4th order in $\ve m$ gives information about the mean-field critical exponent $\beta$ of the transitions. For half-filling ($f=0$) the Ising transition has $\beta=1$. This anomalous exponent is due to the linear energy-dependence of the density of states near the charge neutrality point. For $f\neq 0$, the Ising and XY transitions have $\beta=1/2$, since in this case the density of states is finite at the Fermi level.

I now turn to the question if the above mean-field statements are valid. The effective theory for the surface spin-polarization $S_\mu = \sum_{\ve k,\sigma,\sigma'} c^\dagger_{\ve k\sigma} s^\mu_{\sigma\sigma'}c_{\ve k\sigma'}$ in the XY phase is $H^{xy}_{\rm eff} = D S_z^2$, with $D$ a positive constant (see Ref. \onlinecite{superspin}), and $\langle\ve S^2\rangle >0$ for $U>U_{c,xy}$. The mean-field approach in the present work corresponds to treating $\ve S$ as a classical vector, which may point in any in-plane direction in the XY phase. A simple quantum-mechanical treatment of $H_{\rm eff}^{xy}$, however, gives rise to a non-degenerate ground state and zero spin-polarization $\langle S_\mu\rangle = 0$ ($\mu=x,y,z$), but with $\langle\ve S^2\rangle\neq0$ \cite{superspin}. In the thermodynamic limit $\langle\ve S^2\rangle\rightarrow\infty$. Thus, the XY phase describes a large superspin pointing in no direction, which is in a sense the least classical scenario. This is an indication that the mean-field theory of the XY phase does not describe all magnetic properties correctly. Nevertheless, small islands at the TI surface, for which mean-field theory predicts an in-plane spin polarization, are expected to show magnetic characteristics similar to the superspins at the ends of zigzag carbon nanotubes \cite{superspin}.

In the Ising phase the effective theory is $H^z_{\rm eff} = D' S_z^2$ with $D'<0$ and the ground state is two-fold degenerate. In this case, the thermodynamic limit corresponds to the classical limit of a large superspin which points in $\pm z$ direction. Furthermore, it is well known that a 2D Ising ferromagnet may develop long range order at finite temperatures. Therefore it can be expected that the Ising phase is thermodynamically stable.

{\it How to generate the cubic terms.} In the standard k-space model Hamiltonian for a 3D TI \cite{natphys_model,liu_model_hamiltonian_2010} there are two relevant types of terms. The term of the form $H_0=\tau_z \mathcal M(\ve k)$, with $\tau_\mu$ the Pauli matrices acting in orbital space and $\mathcal M(\ve k) = M - 2 B(3-\sum_\mu\cos k_\mu)$, is an identity in spin space and is essentially responsible for creating a gap in the band structure. The term of the form $H_{so}=\tau_x s_\mu A_\mu(\ve k)$ is, among other things, responsible for the spectrum of the surface states. Following the spirit of $\ve k\cdot\ve p$ theory, $A_\mu(\ve k) \simeq A k_\mu$ is usually truncated after the linear order, which leads to a linear surface state spectrum. Consequently, only nearest neighbor hoppings are usually taken into account in the corresponding tight-binding models. As will be shown now, an extension beyond nearest neighbor hopping gives rise to more versatility in the surface state spectrum, including the possibility of vanishing linear and finite cubic terms.

\begin{figure}
\centering
\includegraphics[width=220pt]{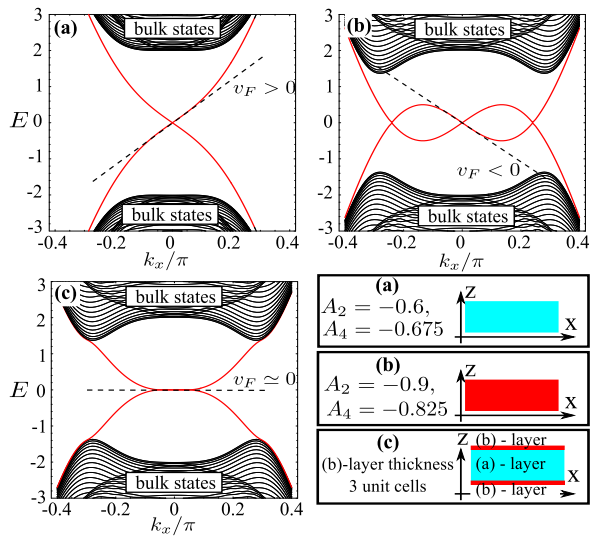}
\caption{(Color online) Non-interacting surface state spectra in homogeneous slabs of 3D TIs for different $A_2$ and $A_4$ parameters [(a) and (b)]. The Fermi velocity of the surface states has opposite signs in (a) and (b). Part (c) shows the surface state spectrum for a layer structure in $z$ direction with a homogeneous $x$ and $y$ direction. The other parameters are equal in all plots: $M = -0.5,\; B=-0.25,\; A_1=3,\; A_3=0$.}
\label{fig_lattice}
\end{figure}

A natural extension of the nearest neighbor SO hopping is
\begin{equation}
H_{so} = \sum_{\boldsymbol \delta, \ve n} A_{|\boldsymbol\delta|} \Psi^\dagger_{\ve n}(\boldsymbol \delta\cdot \ve s) |\boldsymbol\delta|^{-1} \tau_x \Psi_{\ve n+\boldsymbol\delta},\label{so_hopping}
\end{equation}
where $\Psi_{\ve n}$ is a four-component spinor \cite{natphys_model,liu_model_hamiltonian_2010}, $\ve n$ runs over a cubic lattice and $\boldsymbol\delta$ connects a site to its neighbors. For nearest-neighbor SO hopping only, Eq. (\ref{so_hopping}) reduces to the usual $A_\mu(\ve k) = A_1\sin k_\mu$. However, if 1st, 2nd and 4th neighbor SO hopping is taken into account \footnote{Third neighbor hopping is left out here for notational simplicity, but can be taken into account easily.}, then $A_\mu(\ve k) = \sin k_\mu(A_1 + 2 A_2\sum_{\nu\neq \mu} \cos(k_{\nu}))+A_4\sin2k_\mu$. For $A_1 +2A_2+2A_4=0$ and $A_1 -4 A_2+8A_4=  0$ one finds a cubic surface spectrum $A_\mu(\ve k) \propto k_\mu \ve k^2$, with $\ve k=(k_x,k_y)$.

In real materials it is difficult to design the hopping amplitudes, which makes homogeneous TIs with $v_F\simeq 0$ unfeasible. A more realistic approach for tuning $v_F$ to zero is based on attaching a thin layer of a TI with opposite sign of $v_F$ to the surface of a given TI. If the thickness of this layer is chosen properly (layer thicknesses are well controlable with molecular beam epitaxy techniques), the opposite linear terms average to zero and only the cubic terms are left. This is illustrated in Fig. \ref{fig_lattice}, where the numerically calculated non-interacting surface spectra of homogeneous TIs with different material parameters [(a) and (b)] are compared to a layer structure [part (c)].

{\it Conclusion.} A mechanism for tuning surface states of a 3D TI to a strongly correlated regime has been discussed. Cubic terms $\propto k^3$ are introduced in the surface state spectrum in such a way that the Fermi velocity $v_F$ may be tuned to zero while keeping a large bulk energy gap. This limit is characterized by strongly correlated surface states with an effectively non-interacting and topologically non-trivial bulk. The strong surface correlations open a gap in the surface states, even though the topological bulk classification wrongly predicts gapless surface states. A mechanism for generating and tuning the cubic terms in an actual TI has been discussed.

I am grateful for stimulating discussions with C. Honerkamp, T. C. Lang, and S. Wessel.

\bibliography{refs}

\end{document}